\documentclass[letterpaper, 10 pt, conference]{ieeeconf}  

\IEEEoverridecommandlockouts                              

\usepackage{graphicx}      
\usepackage{multirow}
\let\labelindent\relax
\usepackage{enumitem}
\usepackage{tabularx}
\usepackage{placeins}
\usepackage{amsfonts}
\usepackage{amsmath}
\usepackage{accents}
\usepackage{mathrsfs}
\usepackage{amssymb, bm}

\newtheorem{hypp}{\sc Assumption}
\makeatother
      
 \makeatletter

\title{\LARGE \bf
A generic fixed-point iteration-based hierarchical control design: Application to a cryogenic process.
}

\author{Xuan-Huy Pham$^{1,2}$, Mazen Alamir$^{1}$, Fran\c cois Bonne$^{2}$ and Patrick Bonnay$^{2}$
\thanks{$^{1}$University of Grenoble Alpes, Gipsa-lab,(xuan-huy.pham@grenoble-inp.fr)}%
\thanks{$^{2}$ University of Grenoble Alpes, IRIG-DSBT,
 F-38000, Grenoble,France}%
}

\begin{document}

\maketitle
\thispagestyle{empty}
\pagestyle{empty}

\begin{abstract}
This paper presents an extension of a recently proposed hierarchical control framework applied to a cryogenic system. While in the previous work, each sub-system in the decomposition needed to show at least one component of the control input, in the present contribution, this condition is removed enabling a higher flexibility in the definition of the decomposition graph. The impact of this extended flexibility on the computation time is shown using the same cryogenic station where a decomposition in four sub-system is made possible (instead of two in the previous setting). 
\end{abstract}

\section{Introduction}
Cryogenic refrigerators are critical parts of most applications that use the properties of superconducting materials, including nuclear fusion reactors and particle gas pedals: \cite{HENRY20071454,claudet2000economics}. 

In the last decade, system modeling and model-based control method for such systems have been investigated: In (\cite{Bonne_2015}), a library was developed that helps simulating the dynamics of cryogenic systems and for control design preliminary validation; Moreover in \cite{Bonne2014ExperimentalIO} and the references therein, several model-based multi-variable and constrained control strategies have been investigated. However, the aforementioned works are based on centralized frameworks which comes with some obvious drawbacks. Indeed, for such large systems where subsystems might be geographically in different buildings; operators prefer a degree of modularity for easy testing, updating and maintenance operations. On the other hand, PID based completely decentralized schemes fail in achieving optimal design and constraints satisfaction ability.

Recently, a hierarchical control framework has been suggested by \cite{alamir2017fixed}, in which the system under study is decomposed into a network of interacting subsystems. This proposed methodology has been positioned within the huge literature on distributed and hierarchical control strategy \cite{SCATTOLINI2009723,Negenborn2014}. The hierarchical framework is structured in two distinct layers. At the lower local layer is the subsystem layer where each subsystem implements a linear controller in order to regulate a specific output vector. The upper layer consists in a coordinator that exchanges information with the subsystems and optimizes the overall performance by minimizing a global by appropriately choosing a set-points vector that is then sent to the sub-systems (each receives its own set-point vector). In \cite{Pham2021}, the same framework is validated in the presence of model nonlinearities and in the presence of constraints while using a complexity reduction technique to ensure real-time implementability. 

In the two previous works however, the decomposition of the whole process into a network of sub-systems is constrained by the fact that each sub-system is a \textit{controlled system}, i.e.  having at least one control input and one or more regulated outputs. For a large systems, this might be problematic (and it is shown hereafter that for the experimental cryogenic station at hand, it already is when nonlinear MPC is used) since the number of control inputs is generally much smaller than the dimension of the state. Indeed, if a concentrated but strong nonlinearity is present requiring nonlinear MPC to be used, the above mentioned decomposition constraint might imply that whatever is the \textit{admissible decomposition}, the nonlinearity necessarily belongs to a high dimensional subsystem even if the major part of the sub-system show linear dynamics. This results in a high dimensional nonlinear MPC becoming the bottleneck node for the overall computation time. Instead, such newly promoted sub-systems might represent constraint violation or simply coherence constraints between auxiliary signals. 

This is the starting point of the the present contribution. Indeed, the above mentioned constraint on the decomposition graph is relaxed making \textit{eligible} decomposition architectures where some of the sub-systems show no control or even no regulated output. 

The paper is organized as follows: Section \ref{sec-hierachical} describes the hierarchical framework investigated in this paper extending the eligibility condition of a decomposition topology. Section \ref{fixed_sec} briefly recalls the different steps of the overall hierarchical control. Finally, section \ref{sec-sim} presents numerical simulation showing the significant advantaged that can be obtained thanks to the extension of eligible decomposition topology. 

{\bf Notation}. The following notation is extensively used in the paper. For a sequence of vector $q_{i_1}, q_{i_2}, \dots$, the following notation concatenation operator is used:
\begin{equation}
\underset{i \in \mathcal{I}}{\oplus}q_i : = [q_{i_1}^T,q_{i_2}^T, \dots]^T,\, \, \text{with} \quad i_1 < i_2 < \dots \in \mathcal{I}
\end{equation}
Moreover, the bold-faced notation $\bm p$ denotes the profile of a vector variable $p$ over a prediction horizon of length $N$, namely:
\begin{equation}
\bm p = [p^T(k), \dots, p^T(k+N-1)]^T \in \mathbb{R}^{N\cdot n_p}
\end{equation}
\section{Hierarchical control formulation} \label{sec-hierachical}
\begin{figure}[h]
 \centering
 	\includegraphics[]{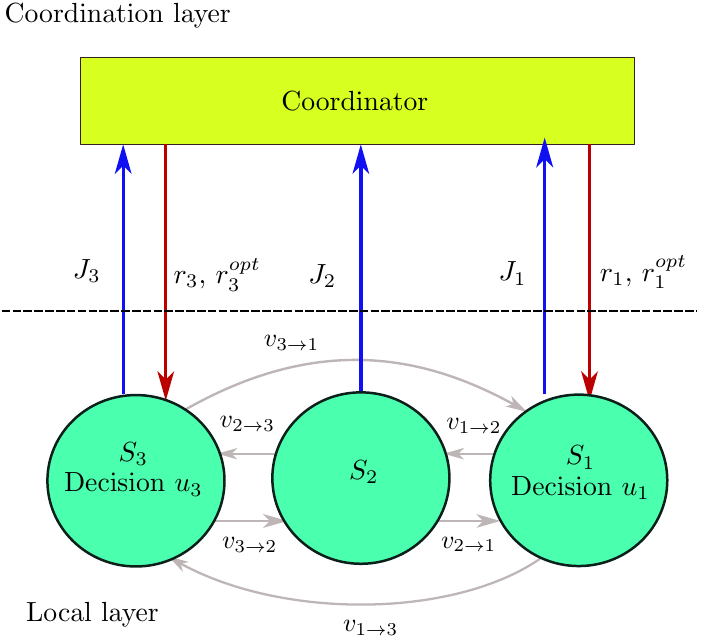}
 	\caption{Example of the hierarchical control architecture and the interconnection network between the subsystems. The presented sets correspond to this example are $\mathcal{N}:= \{1,2,3\}$; $\mathcal{N}^{ctr} := \{1,3\}$, $\mathcal{N}_1 = \{2,3\}$, $\mathcal{N}_2 = \{1,3\}$, $\mathcal{N}_3 = \{1,2\}$.} \label{Fig_inter}
 \end{figure}

In order to better understand the paradigm studied in this contribution, let us consider the situation described in Fig. \ref{Fig_inter} where a set of interacting subsystems indexed by $\mathcal{N}:= \{1,\dots,n_s\}$ is represented. This set is subdivided into two different subsets: 
\begin{itemize}
\item The subset of controlled subsystems $\mathcal N^{ctr}\subset \mathcal N$ having each its control input vector and regulated output vector, denoted by $u_s$ and $y_s$ respectively. 
\item The complementary subset of subsystems that includes no control input denoted by $N^{unc}:= \mathcal N-\mathcal N^{ctr}$.
\end{itemize}  
Each subsystem $S_s$ sees its dynamics impacted through the so-called coupling signale $v_{s' \rightarrow s}$ coming from all subsystems  $\{S_{s'}\}_{s'\in \mathcal{N}_s}$ with indices $s'$ that belong to the set of indices $\mathcal N_s$ (set of indices of subsystems impacting $S_s$).

Let $\bm v^{in}_s$ and $\bm v^{out}_s$ denote respectively the incoming/outgoing coupling profiles of the subsystem $S_s$. More precisely:
\begin{equation}
\bm v_s^{in} :=   \underset{s' \in \mathcal{N}_s}{\oplus} \bm v_{s' \rightarrow s} \quad;\quad 
\bm v_s^{out} := \underset{s' \vert s \in \mathcal{N}_{s'}}{\oplus} \bm v_{s \rightarrow s'}
\end{equation}
The following assumption is considered regarding the processing that is available locally at the subsystem's level:

\vskip 2mm
\hrule 
\vskip 2mm
\begin{hypp}
Each subsystem $S_s$, when given 
\begin{itemize}
\item a presumed incoming profile $\bm v_s^{in}$ and
\item a given individual set-point $r_s$ (required if $s\in \mathcal N^{ctr}$),
\end{itemize}   
can process an algorithm to compute what would be:
\begin{itemize}
\item Its resulting outgoing profile $\bm v_s^{out}$ and 
\item Its contribution $J_s$ to the central cost
\end{itemize}
The central cost is assumed to be of the form:
\begin{equation}
J_c(r, \bm v^{in}) := \sum_{s\in \mathcal N^{ctr}}J_s(r_s, \bm v_s^{in})+\sum_{s\in \mathcal N^{unc}}J_s(\bm v_s^{in})
\end{equation}  
where $r := \underset{s \in \mathcal{N}^{ctr}}{\oplus}r_s$ and $\bm v^{in} := \underset{s\in \mathcal N}{\oplus}\bm v_s^{in}$
\end{hypp}
\vskip 2mm
\hrule 
\vskip 2mm
Note that the computations processed at the local subsystems' level depends on the current states of each subsystem of which the coordinator is unaware. As a results, each time the coordinator sends $(r, \bm v^{in})$, it receives the information that enables to construct the corresponding $\bm v^{out}$, namely there is a map (that depends implicitly on the current state of the system at the time this request is done by the coordinator):
\begin{equation}
\bm v^{out} =\bm g_{out}(r,\bm v^{in}) \label{defdegout}
\end{equation} 
Note that the elements of the outgoing coupling profile $\bm{v}^{out}$ are also those of the profile $\bm{v}^{in}$ but arranged in a different order. Indeed both $\bm v^{in}$ and $\bm v^{out}$ are composed of all the profiles of the form $\bm v_{s\rightarrow s'}$ for $s\in \mathcal N_{s^{'}}$. Therefore, there exists a matrix $G_{in}$ such that :

\begin{equation}
\bm v^{in} = G_{in} \cdot \bm v^{out} \label{defdeGin}
\end{equation}
Injecting \eqref{defdeGin} in \eqref{defdegout} yields:
\begin{equation}
\bm v^{in}=G_{in}\cdot \bm g_{out}(r,\bm v^{in}) \label{coherenceCostraint}
\end{equation} 
Note that this constraint on $\bm v^{in}$, referred to in the sequel as the \textit{coherence constraint}, is satisfied if and only if when the coordinator sends $\bm v^{in}$ to the subsystems, the resulting $\bm v^{out}$ represent the same $\bm v^{in}$ or in other words is compatible with $\bm v^{in}$. Obviously this implicit equation cannot be satisfied at the first guess and need a sort of fixed-point iteration in order to solve it. This is needed because the coordinator has no idea about the equations that hold \textit{inside the subsystem} which links the $\bm v_s^{in}$ to they corresponding $\bm v^{out}_s$. 

The local costs $J_s$ are to be chosen according to criteria associated to each subsystem. For instance, 
for $\{S_s\}_{s\in \mathcal N^{ctr}}$ an output regulation criterion can be used such as:
\begin{equation}
J_s(r_s, \bm v_s^{in})= \sum_{i=0}^{N-1} \| y_s(k+i)-r^d_s \|^2_{Q_c^{(s)}} + \| u_s(k+i) \|^2_{R_c^{(s)}} \label{cost14}
\end{equation}
where $y_s(k+\cdot)$ is the output profile associated to $(r_s,\bm v^{in}_s)$ as explained above while $r^d_s$ is the desired set-point of the subsystem $S_s$. Note that this has to be distinguished from the set-point $r_s$ sent by the coordinator which plays a simple role of auxiliary parameterization. The reason is that the best $r_s, s\in \mathcal N^{ctr}$ that help minimizing the central cost (based on $r^d$) are not necessarily equal to $r_s^d$. The reader can refer to \cite{alamir2017fixed} for more details. 

For $S_s$ with $s\in \mathcal N^{unc}$ that shows no control input, they might involves some outputs to be constrained. In this case, the cost value $J_s$ to be returned to the coordinator might be of the form:
\begin{equation}
J_s(\bm v_s^{in})= \sum_{i=0}^{N-1} \|\max(y_s(k+i)-\overline{y}_s,0)\|^2_{Q_c^{(s)}}\label{cost3}
\end{equation}
where again, $y_s(k+i)$ refers to the trajectory of the subsystem should the incoming signal be $\bm v^{in}_s$ while $\overline{y}_s$ stands for the upper bound on the output $y_s$.

Finally, we have all we need to state the central optimization problem at the coordination layer, namely:
\begin{align}
&\min_{r, \bm v^{in}} \quad J_c(r, \bm v^{in}) \label{central_cost} \\
&\text{subject to} \quad  \bm v^{in}=G_{in}\cdot \bm g_{out}(r,\bm v^{in}) \label{cohconstr}
\end{align}
As a matter of fact, the true decision variable in the above optimization problem is $r$ since the constrains \eqref{cohconstr} fully determine $\bm v^{in}$ for any $r$. The difficulty lies in the fact that the mathematical expression involved in \eqref{cohconstr} is totally unknown to the coordinator. That is the reason why, in order to solve the optimization problem, \cite{alamir2017fixed} proposed an algorithm based on fixed-point iteration, in which, for a frozen auxiliary set-point $r$ sent by the coordinator, the coordinator and the subsystems exchange estimates of $\bm v_s^{out,(\sigma)}$ and $\bm v^{in,(\sigma)}_s$ until the coherence constraint \eqref{cohconstr} is satisfied leading to the corresponding value of the incoming sequence $\bm v^{in,\star}(r)$. This provides the coordinator with the estimation of the central cost for this specific value of the auxiliary set-point $r$, namely $J_c(r, \bm v^{in,\star}(r))$. 

Repeating this process for different values, the coordinator disposes of a successive clouds of values of the form: \vskip 1mm
\begin{equation}
\Bigl\{r^{(i)},J_c(r, \bm v^{in,\star}(r^{(i)}))\Bigr\}
\end{equation}
\vskip 1mm
That can be used to derive an iterative and model-free solution of the original problem \eqref{central_cost}-\eqref{cohconstr}. (see \cite{alamir2017fixed,Pham2021} for the detailed description of the derivative-free trust region based optimization process) and gets the sub-optimal solution $r^{opt}$ in terms of the auxiliary reference vector $r$. 

\section{Recall on the hierarchical control framework} \label{fixed_sec}
\subsection{Evaluate central cost by using fixed-point iteration:}
In this section, the auxiliary set-point $r$ is assumed to be frozen in some sampling period $[k,k+1]$.  The following describe the fixed-point iteration leading to the computation of $J_c(r, \bm v^{in,\star}(r))$ mentioned briefly in the previous section. The coordinator starts with some initial guesses about the incoming coupling profiles:
\begin{equation}
\bm v_s^{in,(\sigma)}, \quad s \in \mathcal{N}, \quad \sigma = 0
\end{equation}
These current guesses are sent to the subsystems $S_s$, $s \in \mathcal{N}$, so that each subsystem can compute the corresponding outgoing coupling profile $\hat{\bm v}^{out,(\sigma +1)}_s$ and sends it to the coordinator. The value of $\hat{\bm v}^{out,(\sigma +1)}_s$ depends obviously on the current state at the subsystem $S_s$. 

Therefore, the coordinator can allocate the elements of the outgoing coupling profile $\hat{\bm v}^{out,(\sigma +1)}$ in the estimate of the incoming coupling profile of each subsystem that are compatible with the return outgoing profiles:
\begin{equation}
\hat{\bm  v}^{in,(\sigma +1)}_s =  G^{(s)}_{in} \cdot \hat{\bm v}^{out,(\sigma +1)}
\end{equation}
concatenating these profiles enables to update the total incoming profile $\bm v^{in, (\sigma+1)}$ according to a filtering step:
\begin{equation}
\bm v^{in,(\sigma+1)} = (\mathbb{I} - \Pi)\cdot \bm v^{in,(\sigma )} + \Pi \cdot \hat{\bm v}^{in,(\sigma +1)}   \label{filter_eq}
\end{equation}
Then, the coordinator sends $\bm v_s^{in,(\sigma +1)}$ (for $s\in \{1,\dots,n_s\}$) to the subsystem $S_s$ for the next round. Considering the synthesis of the matrix $\Pi$, the reader can consult in the paper of \cite{alamir2017fixed}. Shortly speaking, this filter matrix is computed so that the resulting operator is a contraction. 

\begin{figure}[h]
 \centering
%
%
 	\includegraphics[]{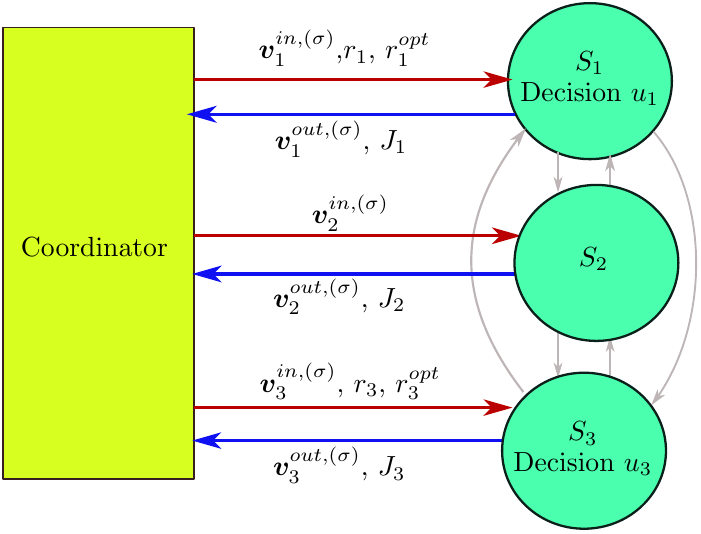}
 	\caption{Schematic of the fixed-point iteration for a given auxiliary set-point $r$.} \label{hier_com}
 \end{figure}

It is essential to note that the coherence constraints \eqref{cohconstr} is fulfilled only when the fixed-point iteration converges to some fixed-point $\bm v^{in,(\infty)}$. In practice, the procedure described above is repeated until some termination condition is met. This can be defined by the logical condition $$\epsilon = \|\bm v^{in,(\sigma+1)}- \bm v^{in,(\sigma)}\| \leq \epsilon_{max}\quad \text{or}\quad \sigma \geq \sigma_{max}$$ 
After the convergence of the iterations, each subsystem $S_s$ computes the the local cost: $$J_s(r|\cdot,\bm v^{in,(\infty)})$$ and sends it back to coordinator in order to compute the central cost, namely:
\begin{equation}
J_c(r|\bm v^{in,(\infty)}) = \sum_{s \in \mathcal{N}} J_s(r,\bm v^{in,(\infty)})
\end{equation}
To conclude, Fig. \ref{hier_com} sketches the whole scheme  of the fixed-point-like round of iterations.
The resolution of $r_s^{opt}$, for $s \in \mathcal{N}^{ctr}$, is described in the next subsection.

\subsection{Optimizing the central cost}
The previous section presented an algorithm that is used to evaluate the central cost $J_c$ associated with a given setpoint $r$. To find the optimal se-tpoint $r^{opt}$, one can use any derivative-free solver  (BOBYQA \cite{Powell2009TheBA}, Genetic algorithm \cite{thede2004introduction}, see also \cite{Cartis2019}). For instance, a method was proposed in \cite{alamir2017fixed} to find the optimal set-point by evaluating each set-point in a $\textbf{G}$ grid. This grid is defined in an iteratively updated trust region built around the previous solution $r^{opt}(k-1)$. By having a map of these set-points, the coordinator can perform a quadratic approximation, which is solved for the optimal set-point. This is also the method used in this paper.

Finally, having the optimal set-point $r^{opt}$ sent from the coordinator, the subsystem $S_s$, $s \in \mathcal{N}^{ctr}$ compute their decision profiles and applies the first element to the plant.
\section{Simulation-based validation} \label{sec-sim}
While the validation of the relevance of the hierarchical framework has been validated in previous works \cite{alamir2017fixed,Pham2021}, the objective of this section is to validate the relevance of extending the eligibility of subsystems definition in the decomposition architecture. The plant is first recalled for which two possible decomposition  topology are defined where the first respect the previous decomposition condition while the second exploits the new degrees of freedom. Comparison between the results of the two architectures in terms of performance and computation time is done in order to validate the contribution of the paper. 
\subsection{Investigated system introduction} 
\begin{figure}[h]
 \centering
%
%
%
%
%
%
 	\includegraphics[]{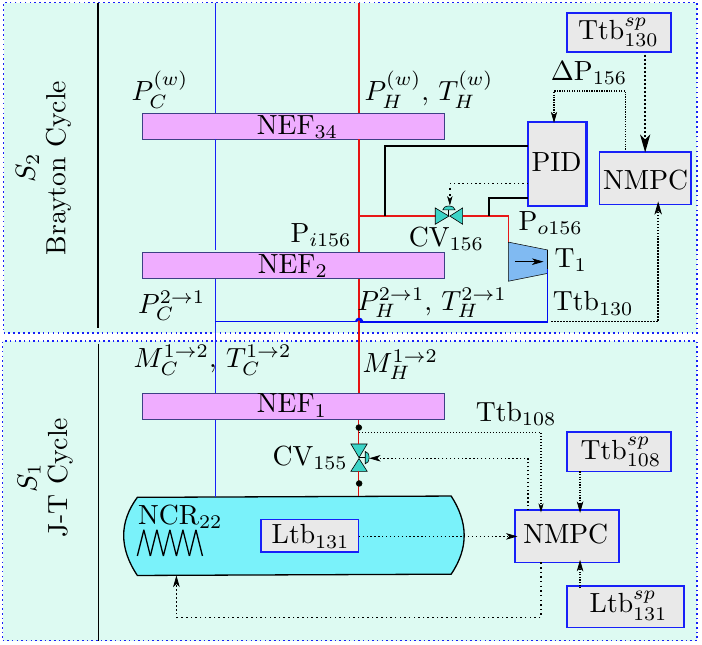}

 	\caption{Block diagram of the cold box plant.} \label{a}
 \end{figure}

The investigated system that are used in order to validate the framework is a cold box of a cryogenic refrigerator. Fig. \ref{a} shows a block diagram of the cold box system consisting of a Joule-Thomson cycle and a Brayton cycle. The Brayton cycle consists of two heat exchangers, which are NEF$_2$, NEF$_{34}$ and a turbine T$_1$. The helium flow is cooled down using the cryogenic turbine T$_1$ to extract thermal energy from the fluid and by exchanging the heat power through a series of heat exchangers (NEF$_x$). A part of the helium gas is liquefied by the valve CV$_{155}$ through the isenthalpic process. The allowed flow rate $M_{out}$ for this installation is $70\ g/s$ for safety reasons.\\
 
\textbf{The Manipulated Inputs:} There are three control inputs which are CV$_{155}$, NCR$_{22}$ belonging to Joule-Thomson cycle and $\Delta \text{P}_{156}$ which is a part of the Brayton cycle. These actuators are defined below:
\begin{enumerate}
\item CV$_{155} \in [0\%,100\%]$: This valve is situated at the inlet of the helium bath. 
\item NCR$^{(a)}_{22}$: This heating actuator is located inside the helium bath ($S_1$). The value of $\text{NCR}_{22}^{(a)}$ is in the range of $[0,55]$ W. Note that the variable NCR$_{22}$ in Fig. \ref{a} is decomposed into two terms:
\begin{equation}
\text{NCR}_{22} := \text{NCR}_{22}^{(a)} + \text{NCR}_{22}^{(w)}
\end{equation}
where $\text{NCR}_{22}^{(w)}$ represents the disturbance coming from the heat source.
\item $\Delta \text{P}_{156} \in  [0,12]$ bar: The pressure drop between the inlet pressure and outlet pressure of the valve CV$_{156}$. It should be noted that the valve CV$_{156}$ is used to control the pressure drop $\Delta \text{P}_{156}$ between its inlet and outlet pressure. To do so, the local NMPC  of $S_2$ computes and sends an appropriate value of the pressure drop $\Delta\text{P}_{156}$ to the PID controller, which acts on the opening position of the valve CV$_{156}$ (Fig. \ref{a}).
\end{enumerate}
\textbf{The Regulated Outputs:} There are three regulated outputs (see Figure \ref{a}. for the notation):
\begin{enumerate}
\item Ltb$_{131}$: The helium liquid level (\%) that must be controlled to ensure that some thermal charges always remain inside the phase separator (e.g. used to cool super-critical helium at liquid helium temperature to be ready for the final customer) are immersed with liquid helium. The set-point is chosen by the operator. In the usual operation, it is set at Ltb$_{131}^{sp}= 60.5\%$.
\item Ttb$_{108}$: The temperature at the inlet of the J-T valve must be tightly controlled in order to ensure the efficiency of the liquefaction of the helium.
\item Ttb$_{130}$: Since the cryogenic turbine is a sensitive component, the temperature at its outlet must be regulated to avoid the risk of solid droplet forming at the outlet, potentially destructive for the turbine.
\end{enumerate}
%

In this contribution, two decomposition topologies of the overall system are compared:

\textbf{Two-subsystems-decomposition (strategy 1):} This decomposition consist of two subsystems that are the Joule-Thomson cycle ($S_1$) and the Brayton cycle ($S_2$) as already depicted in Fig. \ref{a}. The Joule-Thomson has its own control inputs and the controlled outputs that are respectively (CV$_{155}$ + NCR$_{22}^{(a)}$) and (Ltb$_{131}$ + Ttb$_{108}$), while the Brayton cycle has $\Delta$P$_{156}$ and Ttb$_{131}$ as control input and output. The Joule-Thomson cycle is controlled by using Model predictive control (MPC), while the nonlinear MPC (NMPC) is dedicated to controlling the Brayton cycle. Table \ref{table1} summarizes the inputs and outputs of each subsystems in this decomposition. The notations $T_C$, $M_C$, and $P_C$ ($T_H$, $M_H$, and $P_H$) are respectively the temperature, flow rate, and pressure of the cold(hot) branch of the refrigerator.

\begin{figure}[h]
 \centering
 	\includegraphics[]{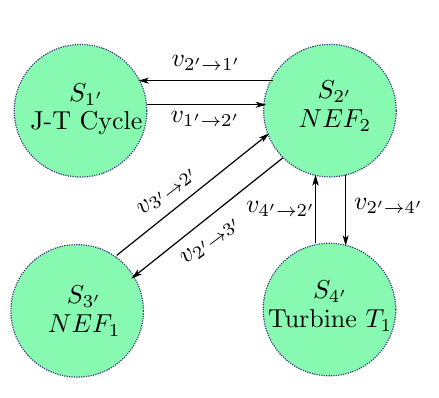}

 	\caption{The interconnection network between the subsystems of the Cold Box.} \label{4ss_inter}
 \end{figure}

\textbf{Four-subsystems-decomposition (strategy 2):} This second configuration consists in decomposing the Brayton cycle into three subsystems, namely the two heat exchangers and the turbine, while keeping the J-T cycle as a subsystem.  In this decomposition, nonlinearity is used in the turbine model while the heat exchanger models are linearized around an operating point $x_{op}$. Note that only the turbine and Joule-Thomson cycle are controlled by NMPC and MPC respectively, while the other subsystems are impacted by their decisions. However, because of the use of only two subsystem in the previous decomposition, the nonlinearity of the turbine makes the whole model nonlinear despite the fact that the heat exchanger part of the model is linear. This time, the labels $S_{x'}$ are taken to represent the subsystems as illustrated in Fig. \ref{4ss_inter}. Table \ref{table_network} shows the intputs, outputs and the coupling signal of this topology.
\begin{table}[h]
\centering
\caption{The inputs, outputs and the coupling variables of two subsystem in 2-subsystems topology} \label{table1}
\renewcommand*{\arraystretch}{1.3}
\begin{tabular}{ccccc}
\cline{2-4}
\multicolumn{1}{c|}{}      & \multicolumn{1}{c|}{$u_s$}                                                             & \multicolumn{1}{c|}{$y_s$}                                                             & $v_{s\rightarrow s'}$                                                                  &  \\ \cline{1-4}
\multicolumn{1}{l|}{$S_1$} & \multicolumn{1}{l|}{\begin{tabular}[c]{@{}l@{}}NCR$_{22}^{(a)}$\\ CV$_{155}$\end{tabular}} & \multicolumn{1}{l|}{\begin{tabular}[c]{@{}l@{}}Ltb$_{131}$\\ Ttb$_{108}$\end{tabular}} & $v_{1 \rightarrow 2} = [M_H^{1 \rightarrow 2}, M_C^{1 \rightarrow 2}, T_C^{1 \rightarrow 2}]^T$ &  \\ \cline{1-4}
\multicolumn{1}{l|}{$S_2$} & \multicolumn{1}{l|}{$\Delta$P$_{156}$}                                                  & \multicolumn{1}{l|}{\begin{tabular}[c]{@{}l@{}}Ttb$_{130}$\\$M_{out}$\end{tabular}}                                                       & $v_{2 \rightarrow 1} = [P_H^{2 \rightarrow 1}, P_C^{2 \rightarrow 1}, T_H^{2 \rightarrow 1}]^T$                                                             &  \\ \cline{1-4}
                           &                                                                                        &                                                                                        &                                                                                        & 
\end{tabular}
\end{table}

\begin{table}[h]
\centering
\caption{The inputs, outputs and the coupling variables of the 4-subsystems topology.} \label{table_network}
\renewcommand*{\arraystretch}{1.3}
\begin{tabular}{c|c|c|l}
\hline
       &   $u_s$      & $y_s$    & \multicolumn{1}{c}{$v_{s \rightarrow s'}$}     \\ \hline                                                                                                                                                                                          
$S_{1'}$  &\begin{tabular}[c]{@{}c@{}}NCR$_{22}^{(a)}$\\ CV$_{155}$\end{tabular}   & \begin{tabular}[c]{@{}c@{}}Ltb$_{131}$\\ Ttb$_{108}$\end{tabular} & $v_{1' \rightarrow 2'} = [M_{H}^{1' \rightarrow 2'},M_{C}^{1' \rightarrow 2'},T_{C} ^{1' \rightarrow 2'}]^T$              \\ \hline
$S_{2'}$       &  \_          & \_     & \begin{tabular}[c]{@{}l@{}l@{}} $v_{2' \rightarrow 1'} = [T_{H}^{2' \rightarrow 1'}, P_{H}^{2' \rightarrow 1'}, P_{C}^{2' \rightarrow 1'}]^T$ \\  $v_{2' \rightarrow 3'} = [M_H^{2' \rightarrow 3'}, M_C^{2' \rightarrow 3'}, T_C^{2' \rightarrow 3'}]^T$ \\  $v_{2' \rightarrow 4'} = [P_C^{2' \rightarrow 4'}]$ \end{tabular} \\ \hline
$S_{3'}$       &\_             & \_                                                                 & \begin{tabular}[c]{@{}l@{}}$v_{3' \rightarrow 2'} = [T_{H}^{3 \rightarrow 2}, P_H^{3' \rightarrow 2'}, P_C^{3' \rightarrow 2'}]^T$\\ $v_{3' \rightarrow 4'} = [T_{H}^{3' \rightarrow 4'},P_H^{3' \rightarrow 4'}]^T$\end{tabular}            \\ \hline
$S_{4'}$    &     $\Delta$P$_{156}$         &\begin{tabular}[c]{@{}c@{}} Ttb$_{130}$ \\ $M_{out}$\end{tabular}                                                      & \begin{tabular}[c]{@{}l@{}}$v_{4' \rightarrow 2'} = [M_C^{4' \rightarrow 2'}, T_C^{4' \rightarrow 2'}]^T$\\ $v_{4' \rightarrow 3'} = [M_H^{4' \rightarrow 3'}]$\end{tabular}                                                             \\ \hline
\end{tabular}
\end{table}

The following local costs of each subsystem are used:

For $S_{1'}$ and $S_{4'}$ that need to track the desired set-point $r^d_s$:
\begin{equation}
J_s(r|r^d_s)= \sum_{i=0}^{N-1} \| y_s(k+i)-r^d_s \|^2_{Q_c^{(s)}} + \| u_s(k+i) \|^2_{R_c^{(s)}} \label{cost14}
\end{equation}
For $S_{3'}$ that has output to be constrained
\begin{equation}
J_{3'}(r|\overline{y}_3)= \sum_{i=0}^{N-1} \|\max(y_{3'}(k+i)-\overline{y}_{3'},0)\|^2_{Q_c^{(3')}}\label{cost3}
\end{equation}
Finally, $S_{2'}$ does not have any contribution to the central cost, its cost is simply defined by $J_{2'}(r) = 0$. The weighing matrices appears in \eqref{cost14}-\eqref{cost3} is listed below:
\begin{align}
Q_c^{(1')} &= \text{diag}([10^4,10^4]) &&\quad Q_c^{(4')} = 10^6  & Q_c^{(3')} &= 10^{12}\\
R_c^{(1')} &= \text{diag}([0,0]) &&\quad R_c^{(4')} = 0
\end{align}
where $\text{diag}()$ denotes a diagonal matrix. 

The local costs for the 2-subsystems topology can simply be deduced from the previous choices so that the resulting central cost is identical. 

In order to compare the performance of two strategies, an index that is necessary is the closed-loop performance $J_{c}^{cl}$, namely:
\begin{align}
J_{c}^{cl} = \frac{1}{N_{sim}} \sum_{i=1}^{N_{sim}} \sum_{s \in \mathcal{N}} J_s^{cl}(i) \\
\end{align}
where $N_{sim}$ is the current simulation time, $J_s^{cl}(i)$ is computed according to the criteria of each subsystem:
\begin{itemize}
\item for tracking set-point:
\begin{equation}
J^{cl}_s(i) = \|y(i) - r^d \|_{Q_c^{(s)}} + \| u_s(i)\|^2_{R_c^{(s)}}
\end{equation}
\item for constraint violation:
\begin{equation}
J^{cl}_s(i) = \|\max(y_s(i)-\overline{y}_3,0) \|_{Q_c^{(s)}}
\end{equation}
\end{itemize}
\subsection{Numerical simulation} 
\begin{figure}[h]
 \centering
  \includegraphics[trim= {1.5cm 5.5cm 1.5cm 1.25cm},width =\linewidth]{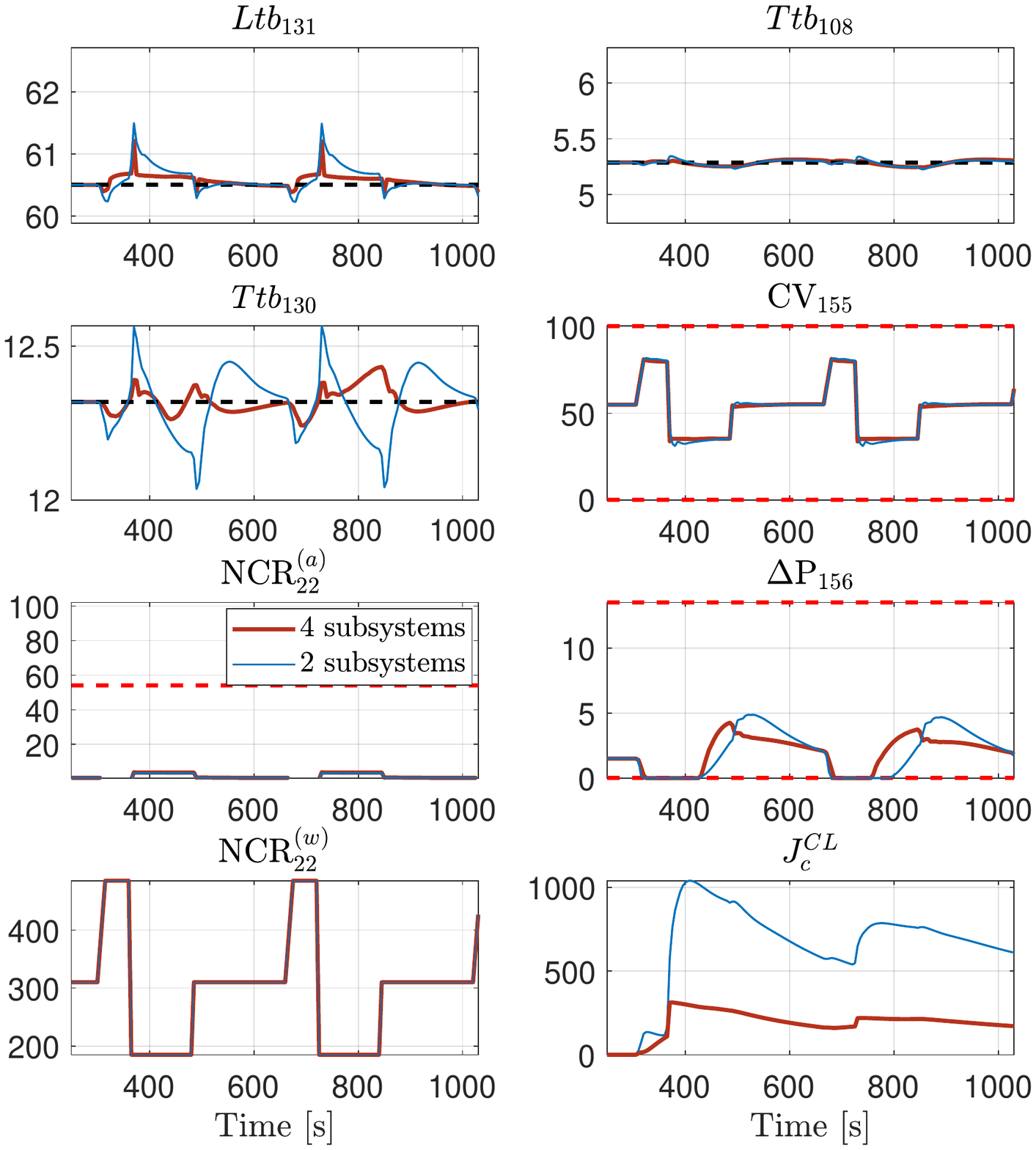} 

  \caption{Comparison of the system behavior between two strategies using hierarchical control.} \label{compare_sp}  
\end{figure}

\begin{figure}[h]
\centering
\includegraphics[width =0.75\linewidth]{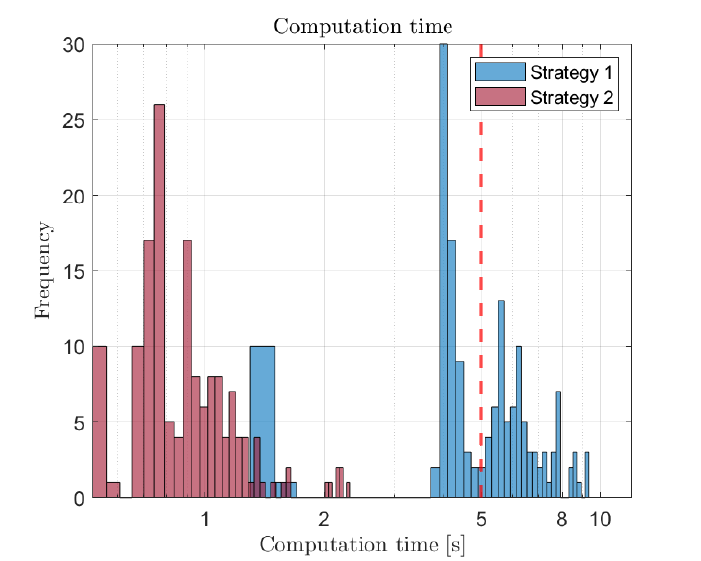}
\caption{The computation time of the NMPC in the two strategies. Note that red dash line depicts the sampling time which is $T_s = 5$ seconds.} \label{cp_time}
 \end{figure}

First, the outcomes of the mentioned strategies, which are the computation time and the closed-loop performance are compared in Fig. \ref{compare_sp} and Fig. \ref{cp_time}, respectively. In Fig. \ref{compare_sp}, the subplot (4,2,8) shows that the 4-subsystems topology is more efficient than the 2-subsystems topology in terms of close-loop performance $J_c^{cl}$. Note also that the computation time of 4-subsystems topology is significantly smaller than the computation time of the 2-subsystems topology since the large-scale problem optimization control problem of $S_2$ (2-subsystems topology) is decomposed into more tractable ones  (Fig. \ref{cp_time}).

\begin{figure}[h]
 \centering
	\includegraphics[width =0.9\linewidth]{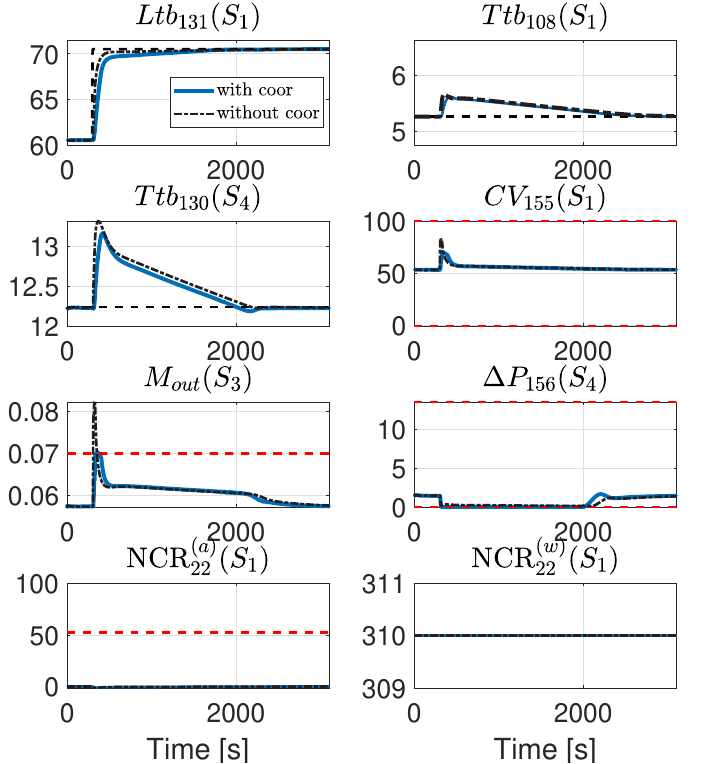}
  \caption{Comparison of the system behavior between using the hierarchical control and observer-based decentralized control. In this scenario, the set-point of liquid helium level Ltb$_{131}$ is changed at instant $t=300$. Under decentralized control, although Ltb$_{131}$ tracks the set-point faster than the one given by hierarchical control, the constraint on $M_{out}$ is violated, which is not the case with the coordination.} \label{compare_cstr}
 \end{figure}

Finally, Fig. \ref{compare_cstr} illustrates the scenario where the upper bound on the flow rate is set equal to $\overline{M}_{out} = 0.07 \, kg/s$. In this scenario, the system behaviors resulted by using decentralized control method (observer-based) and the hierarchical control approach with 4-subsystems topology is compared. Note that in observer-based decentralized control approach, the outgoing coupling signals $v_s^{out}$ are estimated by the observer of each subsystem. The result shows the coordinator drives the system such that the constraint on the flow rate is satisfied.
 Note that the benefit from using hierarchical design compared to decentralized one has been extensively done in \cite{alamir2017fixed,Pham2021} where more dramatic difference has been shown than what the above figure might suggest. 

\section{Conclusion}
In this paper, a hierarchical control method has been applied to a generic case with several subsystem. This approach also allows to take into account the subsystem constraints by defining the corresponding local costs. Promising results with the Simcryogenics library models on a MATLAB simulation were obtained.

Ongoing work aims to validate the control structure, including verifying this approach with a full cryogenic facility (more subsystems) and replacing the filter in \eqref{filter_eq} with a residual-based iterative method so that the coordinator completely ignores knowledge of the subsystem mathematical equations. The application of machine learning to replace the local layer controllers is also considered.
\bibliographystyle{IEEEtran}
\bibliography{IEEEabrv,mybibfile_2}
\end{document}